\pdfoutput=1
\documentclass[twocolumn,
               showpacs,
               preprintnumbers,
               nofootinbib,
               prd,
               10pt,
               notitlepage,
               aps]{revtex4-2}
               
\usepackage{graphicx,amssymb,amsmath,amsthm,amsfonts,epsfig,mathtools}

\usepackage[utf8]{inputenc}
\usepackage{graphicx}
\usepackage{dcolumn}
\usepackage{bm}
\usepackage{color}
\usepackage{soul}
\usepackage[dvipsnames]{xcolor}
\usepackage{hyperref}
\hypersetup{colorlinks=true, citecolor=MidnightBlue,linkcolor=CornflowerBlue, urlcolor=CornflowerBlue, linktocpage=true}
\usepackage{xfrac}
\usepackage{siunitx}

\newcommand{\beqn}{\begin{eqnarray}}
\newcommand{\eeqn}{\end{eqnarray}}
\newcommand{\beq}{\begin{equation}}
\newcommand{\eeq}{\end{equation}}

\def\d{\mathrm{d}}
\def\L{\mathcal{L}}

\def\gbar{\bar{g}}

\begin{document}

\title{Aspects of time evolution in $p$-form field theories}

\begin{abstract}
We show that higher form fields, specifically 2- and 3-forms in four spacetime dimensions, suffer from loss of hyperbolicity when they have self interaction. The equations of motion lose their wave-like characteristic in certain parts of the configuration space, and such problematic regions can also be reached dynamically from healthy initial data. These findings are analogous to recent results for vector fields, which are equivalently 1-forms, however, the details of the pathology can be different for higher forms. We also show that the possibility of more than one self interaction term in higher forms also leads to the amelioration of hyperbolicity problems.
\end{abstract}

\author{K{\i}van\c{c} \.I. \"Unl\"ut\"urk}
\email{kunluturk@ku.edu.tr}
\affiliation{Department of Physics, Ko\c{c} University, \\
Rumelifeneri Yolu, 34450 Sar{\i}yer, \.{I}stanbul, T\"{u}rkiye}

\author{Fethi M. Ramazano\u{g}lu}
\email{framazanoglu@ku.edu.tr}
\affiliation{Department of Physics, Ko\c{c} University, \\
Rumelifeneri Yolu, 34450 Sar{\i}yer, \.{I}stanbul, T\"{u}rkiye}

\date{\today}
\maketitle

\section{Introduction}
$p$-form fields, in other words totally antisymmetric tensor fields, find a wide range of applications in cosmology~\cite{Brown1987, Brown1988, Duff1989, Duncan1990, Bousso2000, Wu2008, Germani2009, Koivisto2009, Koivisto2009a, Koivisto2010, Morais2017, Capanelli2024}, in quantum chromodynamics \cite{Aurilia1979, Ohta1981, Dvali2005, Dvali2006}, and also as gauge fields coupling to extended objects such as strings and branes \cite{Kalb1974, Henneaux1986}. This study aims to investigate pathologies in the time evolution of self-interacting $p$-form fields motivated by the recent findings about vector fields. 

For vector fields, self interaction has been recently shown to cause a breakdown of time evolution in a phenomenon called \emph{loss of hyperbolicity}, rendering the theories unphysical when taken at face value~\cite{EspositoFarese2010, Clough2022, Mou2022, Coates2022, Coates2023a, Coates2023}. We adapt these findings to higher differential forms with a particular emphasis on 2-forms in four dimensions. This is the logical progression of this line of research, since vector fields have a natural identification to 1-forms via raising and lowering of the index. Hence, common properties of $p$-form fields can be used to obtain more widely applicable results.

The underlying reason for the problems of 1-forms (vector fields) is that the equations of motion are not directly governed by the spacetime metric, but by an effective metric which depends on the vector field amplitude. For a range of amplitudes, this effective metric ceases to be Lorentzian. In other words, the originally wave-like equations of motion of the 1-form change into, for example, a Laplace-like form, which cannot be solved as an initial value problem. This so-called loss of hyperbolicity can occur dynamically, where the initially healthy field configurations can break down during the course of their natural time evolution.

We reveal that loss of hyperbolicity is observed in other $p$-form fields as well, making the phenomenon ubiquitous on one hand. However, we also see that the details of \emph{how} loss of hyperbolicity appears can change drastically from the case of 1-forms. For example, the effective metric becomes degenerate in a way that changes the metric signature for 2-form fields, just like the case of 1-forms, however, the standard methods to detect this, e.g. checking the determinant of the effective metric, are not sufficient anymore.

Another radical difference of higher differential forms is that possible self interaction potentials are much more varied compared to those of 1-forms, which also brings variation in how time evolution behaves. For a 1-form, the simplest self interaction term, which is quartic, is unique up to the value of the coupling constant. This is not the case in general, and different possibilities for a 2-form field in four spacetime dimensions lead to distinct outcomes. Moreover, such differences open up the possibility of combining different terms such that problems of hyperbolicity can actually be improved. It turns out that this is indeed the case with 2-forms in four dimensions. Although loss of hyperbolicity still occurs, time evolution is more persistent with such a compound potential compared to theories with simpler self interactions.

This paper is organized as follows. In Sec.~\ref{sec: self-interacting p-form} we examine the hyperbolicity of the $p$-form equations of motion in the presence of the simplest quartic self interaction. We start by summarizing the previous results on 1-forms and then generalize them to higher forms. In Sec.~\ref{sec: loss of hyperbolicity higher forms} we analyze the loss of hyperbolicity in 2-forms and 3-forms. In Sec.~\ref{sec: more general interactions} we consider more general self interaction potentials which are possible for 2-forms in four spacetime dimensions. The existence of such terms allow for an improvement of hyperbolicity, which we examine in Sec.~\ref{sec: improving hyperbolicity}. We provide our summary and conclusions in Sec.~\ref{sec: conclusion}. We provide some of the technical details of our treatment in the appendices, and also discuss the relationship of 3-forms and 1-forms, which leads to a relatively easy way to analyze their dynamics.

Throughout the paper we work with Cartesian coordinates in flat Minkowski spacetime with the mostly plus metric signature and $c=1$. We denote the Hodge dual of a $p$-form $B_{\mu_1 \cdots \mu_p}$ by a tilde;
\begin{equation}
\tilde{B}_{\nu_1 \cdots \nu_{D-p}} = \frac{1}{p!} \epsilon_{\mu_1 \cdots \mu_p \nu_1 \cdots \nu_{D-p}} B^{\mu_1 \cdots \mu_p},
\end{equation}
which is a $(D{-}p)$-form, where $D$ is the number of spacetime dimensions. Antisymmetrization on tensor indices is denoted by square brackets ``[ ]", and indices to be excluded from antisymmetrization are placed between vertical lines, e.g., ``$[\mu|\alpha\beta|\nu\gamma]$" stands for antisymmetrization on $\mu$, $\nu$ and $\gamma$ only. We will keep our notation general, but many of our results are specific to $D=4$. We will explicitly mention whenever this assumption is made.

\section{Self-interacting $p$-form}
\label{sec: self-interacting p-form}

\subsection{Self-interacting 1-form}
\label{sec: self-interacting 1-form}

Let us begin with a summary of loss of hyperbolicity for the 1-form field, which has been studied in some detail. We will mainly follow \textcite{Coates2022}, but additional details can be found in~\cite{Clough2022, Mou2022, Coates2023a, Coates2023, Unluturk2023, Coates2023b, Doneva2024}.

The Lagrangian of a massive, self-interacting 1-form $A$ is given by
\begin{equation}
\label{eq: 1-form Lagrangian}
\L = -\frac{1}{4}F_{\mu\nu}F^{\mu\nu} - \frac{1}{2}m^2 A_\mu A^\mu - V(A),
\end{equation}
where $F_{\mu\nu} = \partial_\mu A_\nu - \partial_\nu A_\mu = 2\partial_{[\mu} A_{\nu]}$ is the \emph{field strength tensor}, $m$ is the mass parameter and $V$ is the self interaction potential.\footnote{ $V$ also depends on the spacetime metric to be strict, which we suppress for brevity.} There is a one-to-one correspondence between the 1-form and a vector field via conjugation with the spacetime metric, $A_\mu = g_{\mu\nu} A^\nu$, hence we will use the two terms interchangeably.

The simplest self interaction term one can construct is the quartic potential $(A_\mu A^\mu)^2$, so that $V$ can be written as
\begin{equation}
\label{eq: 1-form quartic potential}
    V = \frac{\lambda m^2}{4}\left(A_\mu A^\mu\right)^2,
\end{equation}
where $\lambda$ is a coupling constant. The equations of motion that follow from \eqref{eq: 1-form Lagrangian} and \eqref{eq: 1-form quartic potential} read as
\begin{equation}
\label{eq: 1-form eom}
\partial_\mu F^{\mu\nu} = m^2 z A^\nu,
\end{equation}
with
\begin{equation}
\label{eq: 1-form z}
z=1 + \lambda A_\mu A^\mu.
\end{equation}

Applying $\partial_\nu$ to Eq.~\eqref{eq: 1-form eom} and using the antisymmetry of $F^{\mu\nu}$, we see that the equations of motion imply a \emph{generalized Lorenz condition}
\begin{equation}
\label{eq:generalized_lorenz_1form}
\partial_\mu(z A^\mu) = 0.
\end{equation}
This condition can be inserted back into Eq.~\eqref{eq: 1-form eom} to put it into a wavelike form:
\begin{equation}
\label{eq: 1-form eom wavelike form}
\square A^\nu + \partial^\nu \left(\frac{1}{z} A^\mu \partial_\mu z \right) = m^2 z A^\nu,
\end{equation}
with $\square = \partial^\mu \partial_\mu$.

To analyze the hyperbolicity of Eq.~\eqref{eq: 1-form eom wavelike form}, one can linearize it around an arbitrary background solution and assume that the background solution is constant~\cite{Coates2022}. Then, if this linearized, frozen-coefficients version is non-hyperbolic, one can conclude that the original equation \eqref{eq: 1-form eom wavelike form} is also non-hyperbolic \cite{Sarbach2012}.

In Fourier space, $A_\mu=\chi_\mu e^{i k_\nu x^\nu}$, the principal part, that is the part with the highest order of derivatives, of this linearized, frozen-coefficients equation reads as $\mathcal{P}(k)^\mu_{\phantom{\mu}\nu} \chi_\mu$, with
\begin{equation}
\mathcal{P}(k)^\mu_{\phantom{\mu}\nu} = k^2 \delta^\mu_\nu + \frac{2\lambda}{z} A^\alpha A^\mu k_\alpha k_\nu
\end{equation}
being the the \emph{principal symbol}. Here, $A^\rho$ is the background field value and $k^2 = k^\alpha k_\alpha$. Then, in the large $k$ limit, the dispersion relation is given by $\det(\mathcal{P}(k)^\mu_{\phantom{\mu}\nu}) = 0$, which leads to
\begin{equation}
\det(\mathcal{P}(k)^\mu_{\phantom{\mu}\nu}) = \left(g_{\mu\nu} k^\mu k^\nu\right)^{D-1} \frac{\left(\gbar_{\alpha\beta} k^\alpha k^\beta\right)}{z},
\end{equation}
where $D$ is the number of spacetime dimensions. Hence, the dispersion relation for $A_\mu$ is governed, beside the spacetime metric $g_{\mu\nu}$, by an \emph{effective metric} $\gbar_{\alpha\beta}$,
\begin{equation}
\label{eq: 1-form effective metric}
\gbar_{\alpha\beta} = zg_{\alpha\beta} + 2\lambda A_\alpha A_\beta.
\end{equation}
We can then analyze the effective metric to examine the hyperbolicity of the equations of motion.

The effective metric becoming non-Lorentzian is a clear sign of loss of hyperbolicity. To see when this happens, we can calculate the determinant $\gbar = \det(\gbar_{\alpha\beta})$ using the matrix determinant lemma
\begin{equation}
\label{eq: sylvester's det theorem}
\det(M+uv^{\text{T}}) = \left(1 + v^{\text{T}}M^{-1}u\right) \det(M).
\end{equation}
This gives
\begin{equation}
\label{eq: 1-form det gbar}
\gbar = - \left(1 + \lambda A_\mu A^\mu \right)^{D-1} \left(1 + 3\lambda A_\mu A^\mu \right).
\end{equation}
Thus, the effective metric becomes non-Lorentzian if $\lambda A_\mu A^\mu \leq -1/3$, since the signature $(-+ \dots+)$ of a Lorentzian metric necessarily leads to a negative determinant. This means, what is normally a hyperbolic wave-like equation of motion loses this property when $\lambda A_\mu A^\mu \leq -1/3$, rendering time evolution impossible. 

The fact that parts of the 1-form field configuration space do not have well-posed time evolution is already interesting on its own, but the problem goes even deeper. It has been shown that problematic field values can also be dynamically achieved if we start from initially healthy 1-form field configurations for which time evolution is possible~\cite{Coates2022, Coates2023a, Coates2023}.

\subsection{Higher forms}
\label{sec:higher_forms_general}
The Lagrangian~\eqref{eq: 1-form Lagrangian} can be generalized into a Lagrangian of a self-interacting, massive $p$-form field $B_{\mu_1\cdots\mu_p}=B_{[\mu_1\cdots\mu_p]}$ as
\begin{equation}
\label{eq: Lagrangian of p-form}
\L = -\frac{1}{2} \frac{1}{\left(p+1\right)!} M^2 -\frac{1}{2} \frac{1}{p!}m^2 B^2 - V(B),
\end{equation}
with
\begin{equation}
    M^2 = M_{\mu_1\cdots\mu_{p+1}}M^{\mu_1\cdots\mu_{p+1}}, \quad
    B^2 = B_{\mu_1\cdots\mu_p}B^{\mu_1\cdots\mu_p},
\end{equation}
where $M=\d B$ is the field strength tensor or the \emph{exterior derivative} of $B$, i.e.,
\begin{equation}
    M_{\mu_1\cdots\mu_{p+1}} = \left(p+1\right) \partial_{[\mu_1} B_{\mu_2\cdots\mu_{p+1}]}.
\end{equation}
The equation of motion that follows from Eq.~\eqref{eq: Lagrangian of p-form} is
\begin{equation}
\label{eq: p-form eom for general potential}
\partial_\alpha M^{\alpha\mu_1\cdots\mu_p} = m^2 B^{\mu_1\cdots\mu_p} + p!\frac{\partial V}{\partial B_{\mu_1\cdots\mu_p}}.
\end{equation}

The simplest self interaction potential one can write down for \emph{any} $p$-form is the generalization of Eq.~\eqref{eq: 1-form quartic potential} to higher forms, which is
\begin{equation}
\label{eq: p-form simple potential}
    V = \frac{\lambda m^2}{4}\left(\frac{1}{p!}B^2\right)^2.
\end{equation}
With this choice, the equation of motion \eqref{eq: p-form eom for general potential} has the same form as that of the 1-form:
\begin{equation}
\label{eq: p-form eom for simplest quartic potential}
\partial_\alpha M^{\alpha\mu_1\cdots\mu_p} = m^2 z B^{\mu_1\cdots\mu_p},
\end{equation}
where this time 
\begin{equation}
\label{eq: p-form z}
z = 1 + \frac{\lambda}{p!} B^2.
\end{equation}
For $p\geq1$, applying $\partial_{\mu_1}$ to Eq.~\eqref{eq: p-form eom for simplest quartic potential} and using the antisymmetry of $M$, we get an analog of the generalized Lorenz condition;
\begin{equation}
\label{eq: p-form Lorenz condition}
\partial_\alpha \left(zB^{\alpha\mu_2\cdots\mu_p}\right)=0.
\end{equation}

Again for $p\geq1$, we can write the field strength tensor as
\begin{equation}
M_{\alpha\mu_1\cdots\mu_p} = \partial_\alpha B_{\mu_1\cdots\mu_p} - p\,\partial_{\left[\mu_1\right|} B_{\alpha\left|\mu_2\cdots\mu_p\right]}.
\end{equation}
Hence, the field equation \eqref{eq: p-form eom for simplest quartic potential} can be written as
\begin{equation}
\square B^{\mu_1\cdots\mu_p} - p\,\partial^{\left[\mu_1\right|}\partial_\alpha B^{\alpha\left|\mu_2\cdots\mu_p\right]} = m^2 z B^{\mu_1\cdots\mu_p}.
\end{equation}
Using Eq.~\eqref{eq: p-form Lorenz condition}, this becomes
\begin{align}
\label{eq: p-form eom wavelike form}
\square B^{\mu_1\cdots\mu_p} +& \frac{p}{z} \partial_\alpha \partial^{\left[\mu_1\right|} z B^{\alpha\left|\mu_2\cdots\mu_p\right]} \nonumber\\
& + \frac{p}{z} \partial_\alpha z \partial^{\left[\mu_1\right|} B^{\alpha\left|\mu_2\cdots\mu_p\right]} = m^2 z B^{\mu_1\cdots\mu_p}.
\end{align}

We can now proceed in the same manner as in Sec.~\ref{sec: self-interacting 1-form} and obtain the principal symbol
\begin{align}
\frac{1}{p!}\mathcal{P}(k)^{\mu_1\cdots\mu_p}_{\phantom{\mu_1\cdots\mu_p} \nu_1\cdots\nu_p} 
& = k^2 \delta^{[\mu_1}_{\phantom{[}\nu_1}\cdots\delta^{\mu_p]}_{\nu_p} \nonumber\\
&+ \frac{2\lambda}{z\left(p-1\right)!} k_\alpha k^{[\mu_1|} B^{\alpha|\mu_2\cdots\mu_p]} B_{\nu_1\cdots\nu_p}.
\end{align}
Due to antisymmetry, $\mathcal{P}$ is essentially a $d{\times}d$ matrix, where $d=\big(\begin{smallmatrix}
D \\
p
\end{smallmatrix}\big)$ is the number of dimensions of the $p$-form space. We can make this explicit by replacing the antisymmetric sets of indices by multi-indices:
\begin{equation}
\mathcal{P}(k)^a_{\phantom{a}b} = k^2 \delta^a_b + \frac{2p\lambda}{z}C^a B_b,
\end{equation}
where $C^a$ stands for $C^{\mu_1\cdots\mu_p} \equiv k_\alpha k^{[\mu_1|} B^{\alpha|\mu_2\cdots\mu_p]}$. The dispersion relation can now be written as $\det(\mathcal{P}^a_{\phantom{a}b})=0$, where
\begin{align}
\det(\mathcal{P}^a_{\phantom{a}b}) & = \frac{1}{z}\left(k^2\right)^{d-1} \nonumber\\
& \quad \times \left(k^2 z +  \frac{2\lambda}{\left(p-1\right)!} k_\alpha k^{\beta} B^{\alpha\mu_2\cdots\mu_p} B_{\beta\mu_2\cdots\mu_p}\right).
\end{align}
Thus, besides the usual modes with dispersion $g_{\alpha\beta} k^\alpha k^\beta{~=~}0$, we also get the dispersion relation $\gbar_{\alpha\beta} k^\alpha k^\beta = 0$, where the \emph{effective metric} $\gbar_{\alpha\beta}$ is given by
\begin{equation}
\label{eq: p-form effective metric}
\gbar_{\alpha\beta} = zg_{\alpha\beta} + \frac{2\lambda}{\left(p-1\right)!} B^{\phantom{\alpha}\mu_2\cdots\mu_p}_{\alpha} B_{\beta\mu_2\cdots\mu_p}.
\end{equation}

We can now analyze if or when this effective metric becomes singular, which signals loss of hyperbolicity. Even though this formula seems to be completely analogous to the one for the 1-forms in Eq.~\eqref{eq: 1-form effective metric}, we will demonstrate that small details lead to important differences.

\section{Loss of hyperbolicity in higher form fields}
\label{sec: loss of hyperbolicity higher forms}
A clear sign of $\gbar_{\alpha\beta}$ becoming non-Lorentzian is when its determinant $\gbar = \det(\gbar_{\alpha\beta})$ becomes positive. We have already calculated $\gbar$ for a 1-form, where we have made use of the matrix determinant lemma. For higher forms, however, the effective metric \eqref{eq: p-form effective metric} does not have the simple form of a matrix plus the outer product of two vectors as in Eq.~\eqref{eq: sylvester's det theorem}. Hence, $\gbar$ does not have a simple form valid in a general number of dimensions. \emph{We shall therefore specialize to $D=4$.}

In four dimensions, the highest form one can have is a 4-form.  However, for $p=4$, the Lagrangian \eqref{eq: Lagrangian of p-form} does not give a dynamical theory, since $M=\d B$ vanishes identically. Therefore, we only need to consider $p=2,3$.

\subsection{2-form fields}
\label{sec: loss of hyperbolicity 2-forms}

For a 2-form, the determinant of Eq.~\eqref{eq: p-form effective metric} is\footnote{For $p=2$, Eq.~\eqref{eq: p-form effective metric} can be factorized as $\gbar_{\alpha\beta} = z g_{\alpha\mu} \big(g^{\mu\nu} + \sqrt{\frac{2\lambda}{z}} B^{\mu\nu}\big) \big(g_{\nu\beta} - \sqrt{\frac{2\lambda}{z}} B_{\nu\beta}\big)$. Taking the determinant, both factors inside the parentheses give the same Born-Infeld type determinant \cite{Born1934}, hence follows Eq.~\eqref{eq: 2-form det gbar}.}
\begin{equation}
\label{eq: 2-form det gbar}
\bar{g}^{(2)} = - \left[z\left(1 + \frac{3\lambda}{2}B_{\mu\nu}B^{\mu\nu}\right) - \left(\frac{\lambda}{2} B_{\mu\nu}\tilde{B}^{\mu\nu}\right)^2 \right]^2.
\end{equation}

There is an important difference between this expression and the determinant of the effective metric for the 1-form~\eqref{eq: 1-form det gbar}. Namely, for a 2-form, there is no part of the configuration space where $\bar{g}^{(2)}$ is positive. Hence, solely based on the determinant, it might still be the case that the field equations for 2-forms are \emph{almost always} hyperbolic. We will see that this is not the case, and hyperbolicity is indeed lost at $\bar{g}^{(2)}=0$.

Let us start with showing that $\bar{g}^{(2)}=0$ is indeed dynamically encountered if we start with healthy initial data. We define $x = (\lambda/2) B_{\mu\nu}B^{\mu\nu}$ and $y = (\lambda/2) B_{\mu\nu} \tilde{B}^{\mu\nu}$, and write Eq.~\eqref{eq: 2-form det gbar} as
\begin{equation}
\bar{g}^{(2)} = - f(x,y)^2,
\end{equation}
with $f(x,y)=3x^2 + 4x + 1 - y^2$. For a given value of $y$, the determinant vanishes if
\begin{equation}
\label{eq: xpm}
x = x_{\pm} = \frac{-2 \pm \sqrt{1+3y^2}}{3}.
\end{equation}
We are interested in the case where the field starts from a configuration where time evolution is healthy. This means $x$ and $y$ are initially not very large, but afterwards grow to reach the singularity of $\gbar_{\alpha\beta}$. Therefore, the first root to be encountered is $x=x_+$. We note that $x_+ \geq -1/3$ for all values of $y$.

In order to see how $x=x_+$ can be attained dynamically, let us introduce the three-vectors
\begin{subequations}
\begin{align}
\mathbf{E} &= (B_{10},B_{20},B_{30}) \\
\mathbf{B} &= (B_{23},B_{31},B_{12})
\end{align}
\end{subequations}
in analogy with electromagnetism. In a slight abuse of terminology, we shall call $\mathbf{E}$ and $\mathbf{B}$ the electric and magnetic parts of $B_{\mu\nu}$, respectively. Note that with this notation we have $x = \lambda(\mathbf{B}^2 - \mathbf{E}^2)$ and $y = 2\lambda \mathbf{E}\cdot\mathbf{B}$.

Similarly setting $\mathbf{M} = (M_{023}, M_{031}, M_{012})$, Eqs.~\eqref{eq: p-form eom for simplest quartic potential} and \eqref{eq: p-form Lorenz condition} can be written as the first order system
\begin{subequations}
\label{eq: time evolution system}
\begin{align}
\dot{\mathbf{B}} & =\mathbf{M} - \nabla\times\mathbf{E}, \\
\dot{\mathbf{M}} & =\nabla\left(\nabla\cdot\mathbf{B}\right)-m^{2}z\mathbf{B}, \\
\mathbb{A}\dot{\mathbf{E}} & = \nabla\times\left(z\mathbf{B}\right) - 2\lambda\left(\mathbf{B}\cdot\dot{\mathbf{B}}\right)\mathbf{E}, 
\end{align}
\end{subequations}
subject to the constraint
\begin{equation}
\label{eq: constraint eqn}
\nabla\times\mathbf{M}+m^{2}z\mathbf{E}=0,
\end{equation}
where an over-dot signifies a time derivative. Here, the matrix $\mathbb{A}$ is
\begin{equation}
\mathbb{A} = z \mathbb{I}_3 - 2\lambda \mathbf{E} \mathbf{E}^{\text{T}},
\end{equation}
where $\mathbb{I}_3$ is the $3{\times}3$ identity matrix, and $\mathbf{E}^{\text{T}}$ denotes the transpose of the column vector $\mathbf{E}$. We remark that Eq.~\eqref{eq: constraint eqn} also implies $\nabla\cdot \left(z \mathbf{E} \right) = 0$.

Let us first consider $\lambda<0$, for which it is not difficult to find a configuration which breaks down dynamically. Take, for example, the initial configuration
\begin{subequations}
\begin{align}
\mathbf{E} &= 0, \\
z\mathbf{B} &= \nabla\psi, \\
\mathbf{M} &= \kappa\nabla\psi,
\end{align}
\end{subequations}
where $\psi$ is some scalar field and $\kappa$ is a positive constant. This configuration satisfies the constraint \eqref{eq: constraint eqn} and gives $y=0$. The evolution equations \eqref{eq: time evolution system} give $\dot{\mathbf{E}}=0$ and $\dot{\mathbf{B}}=\kappa\nabla\psi$. Thus, after a small time $\Delta t$, $\mathbf{E}$ and $y$ remain zero and $\mathbf{B}$ becomes
\begin{equation}
\mathbf{B}(t+\Delta t) = \left(\frac{1}{z} + \kappa\Delta t\right)\nabla\psi.
\end{equation}
Now, we can choose $\psi$ such that $(\nabla\psi)^2 \lesssim 4/27|\lambda|$, so that we can initially have $x \gtrsim -1/3$. It is therefore obvious that for sufficiently large $\kappa$, the system is going to hit $x=x_+$, and the time evolution breaks down.

For $\lambda > 0$, it is not straightforward to find such an initial configuration, as we discuss in Appendix~\ref{apdx: coordinate singularity}. However, based on similar results concerning 1-forms~\cite{Coates2022,Coates2023a}, we expect that loss of hyperbolicity is again encountered.

Recall that we go back to $\bar{g}^{(2)}<0$ after encountering $\bar{g}^{(2)}=0$ momentarily, so we might play devil's advocate and consider this form of loss of hyperbolicity to be a ``mild'' issue. If the timelike direction determined by $\gbar_{\alpha\beta}$ remained the same before and after $\bar{g}^{(2)}=0$, that is the metric signature $({-}{+}{+}{+})$ stayed the same for the same respective spacetime coordinates, one might hope to somehow evolve the system past $\gbar^{(2)}=0$ by following an externally imposed junction condition. However, a more severe scenario is when the timelike direction changes as $\gbar^{(2)}$ crosses zero; $({-}{+}{+}{+})\to({+}{-}{+}{+})$, or when two other directions become timelike as well; $({-}{+}{+}{+})\to({-}{-}{-}{+})$. We shall show that these two pessimistic cases are what happens with 2-forms,\footnote{One might initially consider the possibility that the former is also a mere coordinate singularity as in the case of the Schwarzschild metric in the standard coordinates: $\d s^2 = -(1-2M/r)\d t^2+(1-2M/r)^{-1} \d r^2 + r^2\d\Omega_2^2$. Investigating this in detail would be a lengthy project on its own which we will not attempt here, see Appendix~\ref{apdx: coordinate singularity} for some relevant discussion. However, note that there are other fields in the universe, for example scalars, whose dynamics are directly controlled by the spacetime metric where the spacetime coordinate $t$ already plays the role of time. Changing this coordinate into a spacelike one for the self-interacting 2-form would already render the total physical system ill-posed.} and hyperbolicity is indeed lost as we mentioned.

We can test this in an arbitrary Lorentz frame, where the four eigenvalues of $\gbar_{\alpha\beta}$ can be computed as
\begin{equation}
X \pm \sqrt{X^2 + f(x,y)} \quad \text{and} \quad 2x+1 \pm \sqrt{\left(2x+1\right)^2 - f(x,y)}.
\end{equation}
Here, $X$ denotes\footnote{Eigenvalues of a rank-$(0,2)$ tensor are naturally not Lorentz invariant. However, the \emph{number} of its positive (negative) eigenvalues is, by Sylvester's law of inertia.}
\begin{equation}
X = \lambda\left(\mathbf{B}^2 + \mathbf{E}^2 \right).
\end{equation}
Note that, if $\lambda<0$, then $X<0$. Therefore, when $f(x,y)$ transitions from positive to negative, three of the eigenvalues become negative, i.e., the effective metric transitions to $({-}{-}{-}{+})$ signature. If $\lambda>0$, however, the signature has to stay mostly plus, but the timelike direction changes when $f(x,y)$ becomes negative. This demonstrates that continuing the time evolution beyond $\bar{g}^{(2)}$ is not possible.

\subsection{3-form fields}
\label{sec: loss of hyperbolicity 3-forms}

Dynamical properties of 3-form fields have some radical differences from those of 2-forms in $D=4$, even in the simplest cases. To begin with, the free, massless 3-form theory with $m^2=0=V(B)$ is not dynamical, $B_{\mu\nu\alpha}$ is pure gauge. However, just as in the case of 1-forms, addition of mass breaks gauge symmetry, and this leads to a dynamical theory. A discussion of these subtleties can be found in Appendix~\ref{apdx: 3-form dynamics}. In the following, we will continue with our main interest, the effect of self interaction together with the mass term. This again leads to a dynamical theory, and the question is whether it suffers from loss of hyperbolicity.

We have already calculated the effective metric which controls the dynamics of a massive, self-interacting 3-form in Sec.~\ref{sec:higher_forms_general}. Calculating the determinant of Eq.~\eqref{eq: p-form effective metric} for a 3-form in four dimensions, we get
\begin{equation}
\label{eq: 3-form det}
\bar{g}^{(3)} = - \left(1 + \frac{\lambda}{6} B^2\right) \left(1 + \frac{\lambda}{2} B^2\right)^3.
\end{equation}
This expression can change sign, hence the effective metric does become non-Lorentzian. Therefore, we expect loss of hyperbolicity to be encountered in this theory as well. We will not have a detailed look at specific initial data that can dynamically lead to the breakdown of time evolution, yet, all the examples so far indicate that it is most likely the case here as well.

We want to note that Eq.~\eqref{eq: 3-form det} informs us about the nature of the loss of hyperbolicity for 3-forms in addition to its existence. The determinant of the effective metric does not only ``touch'' the configuration where it becomes singular---the case with the 2-form fields---but it also changes its sign as for the 1-form fields. Thus, we conjecture the time evolution problems to be more similar to the latter case.

\section{More general interactions}
\label{sec: more general interactions}

The potential \eqref{eq: p-form simple potential} is the only possible quartic self interaction for a 1-form, however, for higher forms, other contractions are possible. For example, for $p=3$, quartic contractions other than Eq.~\eqref{eq: p-form simple potential} are possible, namely,
\begin{equation}
B_{\mu\alpha\beta}B^{\mu\alpha\nu}B^{\beta\gamma\sigma}B_{\nu\gamma\sigma} \quad \text{and}\quad B_{\mu\alpha\beta}B^{\mu\nu\gamma}B^{\alpha}_{\phantom{\alpha}\nu\sigma}B^{\beta\sigma}_{\phantom{\beta\sigma}\gamma}.
\end{equation}
However, in $D=4$, these are equivalent to the term in Eq.~\eqref{eq: p-form simple potential}. This should be so, since in four dimensions a 3-form is Hodge dual to a vector, and Eq.~\eqref{eq: p-form simple potential} is the only quartic term possible for a vector.

The situation is nontrivial for 2-forms, where two classes of potentials are possible: those that do not include the Levi-Civita symbol (parity symmetric) and those that do (parity violating). In the following, we shall examine the former, and leave the parity violating terms to Appendix~\ref{apdx: parity violating interactions}.

For a 2-form, a cubic contraction is possible; $B_{\mu\nu} B^{\mu\alpha} B^\nu_{\phantom{\nu}\alpha}$, however, it vanishes identically. Thus, the simplest self interaction for a 2-form is quartic. Aside from \eqref{eq: p-form simple potential}, the only parity symmetric quartic potential is
\begin{equation}
V = \frac{\beta m^2}{8} B_{\mu\nu}B^{\nu\alpha}B_{\alpha\beta}B^{\beta\mu}.
\end{equation}
This gives the equation of motion
\begin{equation}
\label{eq: 2-form equation of motion}
\partial_\alpha M^{\alpha\mu\nu} = m^2 B^{\mu\nu} - \beta m^2 B^{\mu\alpha}B_{\alpha\beta}B^{\beta\nu}.
\end{equation}
Applying $\partial_\mu$ yields the generalized Lorenz condition
\begin{equation}
\partial_\mu B^{\mu\nu} = \beta \partial_\mu\left(B^{\mu\alpha}B_{\alpha\beta}B^{\beta\nu}\right).
\end{equation}
Thus, the principal part of Eq.~\eqref{eq: 2-form equation of motion} can be written as
\begin{equation}
\square B^{\mu\nu} - \beta \left[\partial^\mu\partial_\sigma \left(B^{\sigma\alpha}B_{\alpha\beta}B^{\beta\nu}\right) - \left(\mu\leftrightarrow\nu\right)\right],
\end{equation}
which leads to the principal symbol
\begin{align}
\frac{1}{2}\mathcal{P}(&k)^{\mu\nu}_{\phantom{\mu\nu}\alpha\beta} \chi^{\alpha\beta} = \frac{1}{2}k^2\left(\delta^\mu_\alpha \delta^\nu_\beta - \delta^\nu_\alpha \delta^\mu_\beta\right)\chi^{\alpha\beta} \nonumber\\
& \quad\quad -\beta \Big[k^\mu B^{\sigma\nu} B_{\sigma[\alpha}k_{\beta]} + k^\mu k^\sigma B_{\sigma[\alpha}B_{\beta]}^{\phantom{\beta]}\nu} \nonumber\\
& \quad\quad\quad + B^{\sigma\gamma}k^\mu k_\sigma \delta^\nu_{[\alpha}B_{\beta]\gamma} - \left(\mu\leftrightarrow\nu\right) \Big] \chi^{\alpha\beta}.
\end{align}
Again, this is essentially a $d{\times}d$ matrix $\mathcal{P}^a_{\phantom{a}b}$. Calculating its determinant in four dimensions, where $d=6$, we get
\begin{equation}
\label{eq: det of 2-form principal symbol}
\det(\mathcal{P}^a_{\phantom{a}b}) = \left(k^2\right)^4 \left(Q_{\alpha\beta\mu\nu} k^\alpha k^\beta k^\mu k^\nu \right),
\end{equation}
where the rank-4 tensor $Q_{\alpha\beta\mu\nu}$ depends on the field $B_{\mu\nu}$.

To analyze the hyperbolicity of Eq.~\eqref{eq: 2-form equation of motion}, one can start by analyzing the condition
\begin{equation}
\label{eq: quartic form dispersion relation}
Q_{\alpha\beta\mu\nu} k^\alpha k^\beta k^\mu k^\nu = 0.
\end{equation}
We were not able to find a simple factorization of $Q_{\alpha\beta\mu\nu}$ so as to formulate the problem in terms of an effective metric. Therefore, to gain insight, we shall look at the two special cases of purely electric and purely magnetic fields.

We choose an arbitrary Lorentz frame and write $(k^\mu)=(\omega,\mathbf{k})$. We can also set $\mathbf{k}=(0,0,k_z)$ without loss of generality. In the electric case ($\mathbf{B}=0$), Eq.~\eqref{eq: quartic form dispersion relation} can be solved for $\omega$ to get the two dispersion relations
\begin{equation}
\omega^2 = \frac{1 - \beta E_z^2}{1 - \beta \mathbf{E}^2} k_z^2
\end{equation}
and
\begin{equation}
\omega^2 = \frac{1 - \beta\left(E_x^2 + E_y^2 + 3E_z^2\right)}{1 - 3\beta \mathbf{E}^2} k_z^2.
\end{equation}
We see that for $\beta>0$, problematic modes exist.

In the magnetic case ($\mathbf{E}=0$), we similarly get
\begin{equation}
\omega^2 = \left[1 + \beta\left(B_x^2 + B_y^2\right)\right] k_z^2
\end{equation}
and
\begin{equation}
\omega^2 = \frac{1 + \beta\left(3B_x^2 + 3B_y^2 + B_z^2\right)}{1 + \beta \mathbf{B}^2} k_z^2.
\end{equation}
In this case, problematic modes exist for $\beta<0$.

Note that this is the same kind of behavior we saw in Sec.~\ref{sec: loss of hyperbolicity 2-forms}. Namely, for $y=0$, $x_+$ is negative, therefore the equations of motion can break down for $\lambda>0$ ($\lambda<0$) in the purely electric (magnetic) case. This fact can be used to potentially ``improve'' the hyperbolicity of the equations of motion as we shall see in Sec.~\ref{sec: improving hyperbolicity}.

\section{Improving hyperbolicity}
\label{sec: improving hyperbolicity}

We saw that the two possible quartic self interaction terms for the 2-form fields have the same type of behavior, at least when restricted to purely electric and purely magnetic cases: the equations of motion are not hyperbolic after a point if the coupling constant is positive (negative) in the electric (magnetic) case. This suggests that if the two couplings are used together but with opposite signs, the equations may possibly be improved in terms of hyperbolicity, if not altogether saved.

Consider such a compound self interaction potential
\begin{equation}
\label{eq: compound potential}
V = \frac{\lambda m^2}{4} \left(\frac{1}{2} B_{\mu\nu}B^{\mu\nu}\right)^2 - \frac{\beta m^2}{8} B_{\mu\nu} B^{\nu\alpha} B_{\alpha\beta} B^{\beta\mu},
\end{equation}
for which the equations of motion read as
\begin{equation}
\partial_\alpha M^{\alpha\mu\nu} = m^2 z B^{\mu\nu} + \beta m^2 B^{\mu\alpha}B_{\alpha\beta}B^{\beta\nu}.
\end{equation}
Here, $z = 1 + (\lambda / 2) B^2$ is defined the same way as before. Hence, the generalized Lorenz condition becomes
\begin{equation}
\partial_\mu B^{\mu\nu} = -\frac{\beta}{z} \partial_\mu \left(B^{\mu\alpha} B_{\alpha\beta} B^{\beta\nu}\right) - \frac{\lambda}{z} \partial_\mu B_{\alpha\beta} B^{\alpha\beta} B^{\mu\nu}.
\end{equation}
Rewriting the principal part using this condition, and going to the Fourier space, we get the principal symbol
\begin{align}
\mathcal{P}_{\phantom{\mu\nu}\alpha\beta}^{\mu\nu} &= k^{2}\delta_{\alpha}^{\mu}\delta_{\beta}^{\nu} + \frac{2\beta}{z} k^{\mu} \bigg[B_{[\alpha}^{\phantom{[\alpha}\sigma}k_{\beta]}B_{\phantom{\nu}\sigma}^{\nu} + \bigg(\delta_{[\alpha}^{\nu}B_{\beta]\gamma}B^{\sigma\gamma} \nonumber\\
&\quad\quad\quad + B_{\phantom{\sigma}[\alpha}^{\sigma} B_{\beta]}^{\phantom{\beta]}\nu}+\frac{\lambda}{\beta}B_{\alpha\beta}B^{\sigma\nu}\bigg)k_{\sigma}\bigg]-\left(\mu\leftrightarrow\nu\right).
\end{align}

Calculating $\det(P^a_{\phantom{a}b})$, we see that it has the same form as Eq.~\eqref{eq: det of 2-form principal symbol}, for a new $Q_{\alpha\beta\mu\nu}$. We again write $(k^\mu) = (\omega,0,0,k_z)$ and solve the condition $Q_{\alpha\beta\mu\nu}k^\alpha k^\beta k^\mu k^\nu = 0$ for $\omega$. In the electric case we get the two dispersion relations
\begin{equation}
\omega^2 = \frac{1 - \lambda\left(E_x^2 + E_y^2\right) - \left(\lambda-\beta\right)E_z^2}{1 - \left(\lambda-\beta\right)\mathbf{E}^2} k_z^2
\end{equation}
and
\begin{equation}
\omega^2 = \frac{1 - \left(\lambda-\beta\right)\left(E_x^2 + E_y^2 + 3E_z^2\right)}{1 - 3\left(\lambda-\beta\right)\mathbf{E}^2} k_z^2.
\end{equation}
In the magnetic case we get
\begin{equation}
\omega^2 = \frac{1 + \lambda B_z^2 + \left(\lambda-\beta\right)\left(B_x^2+B_y^2\right)}{1 + \lambda\mathbf{B}^2} k_z^2
\end{equation}
and
\begin{equation}
\omega^2 = \frac{1 + \left(\lambda-\beta\right)\left(3B_x^2 + 3B_y^2 + B_z^2\right)}{1 + \left(\lambda-\beta\right)\mathbf{B}^2} k_z^2.
\end{equation}
Thus, the dispersion relations simplify and indeed improve if we set $\lambda=\beta$.

\begin{figure}
    \centering
    \includegraphics[width=\columnwidth]{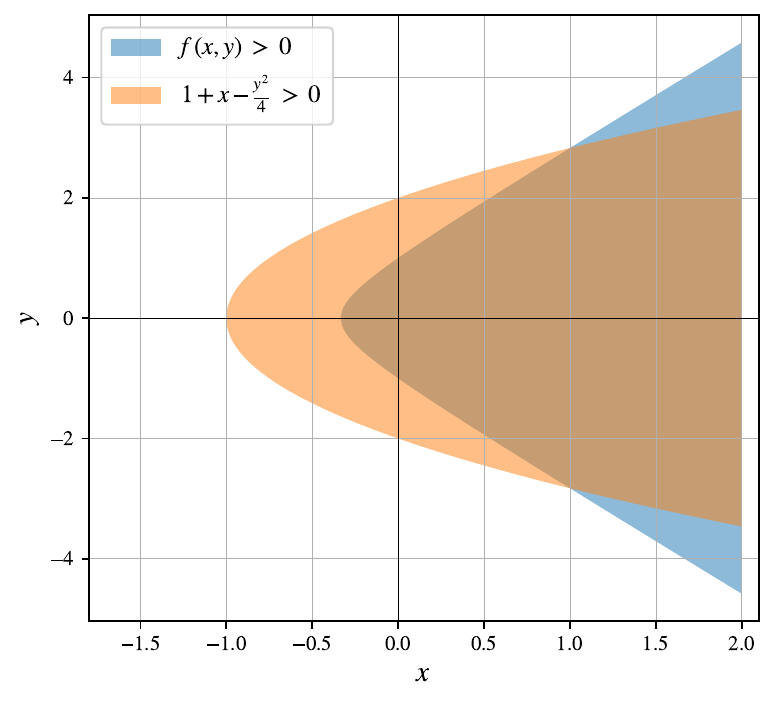}
    \caption{Healthy parts of the configuration space, shown on the $x-y$ plane. Blue region is the healthy part for the simple potential \eqref{eq: p-form simple potential}, orange is for the compound potential \eqref{eq: compound potential}. For small field amplitudes (near the origin), the compound potential significantly enlarges the healthy region, i.e., the loss of hyperbolicity is encountered later. This trend is reversed for $x>1$ or $y^2>8$.}
    \label{fig: healthy regions}
\end{figure}

In fact, for $\lambda=\beta$ the potential \eqref{eq: compound potential} becomes
\begin{align}
V &= \frac{\lambda m^2}{4} \Bigg[ \left( \frac{1}{2} B_{\mu\nu}B^{\mu\nu}\right)^2 
- \frac{1}{2} B_{\mu\nu} B^{\nu\alpha} B_{\alpha\beta} B^{\beta\mu} \Bigg] \nonumber \\
&= - \frac{\lambda m^2}{8} \left(\frac{1}{2} B_{\mu\nu} \tilde{B}^{\mu\nu} \right)^2 .
\end{align}
The determinant of $P^a_{\phantom{a}b}$ can then be calculated explicitly as
\begin{equation}
\det(P^a_{\phantom{a}b}) = \frac{\left(k^2\right)^5}{z^3} \left(\bar{g}_{\alpha\beta} k^\alpha k^\beta\right),
\end{equation}
where the effective metric is
\begin{equation}
\bar{g}_{\alpha\beta} = z g_{\alpha\beta} - \lambda B_\alpha^{\phantom{\alpha}\mu}B_{\beta\mu}.
\end{equation}
The determinant of the effective metric reads in this case as
\begin{equation}
\gbar = - \left(1 + x - \frac{y^2}{4}\right)^2.
\end{equation}
Hence, the singularity is located at $x=-1+y^2/4$. Comparing with $x_+$ in Eq.~\eqref{eq: xpm}, we see that for small values of $y$ ($y^2<8$), the allowed range of $x$ for a given value of $y$ is larger, see Fig.~\ref{fig: healthy regions}. In other words, even though loss of hyperbolicity still occurs, it is encountered later compared to the case where we only have the ``standard'' self interaction term~\eqref{eq: p-form simple potential}.

\section{Conclusion}
\label{sec: conclusion}

The field equations of a self-interacting vector, i.e. 1-form, field can dynamically lose hyperbolicity, which means further time evolution becomes impossible and the theory is rendered unphysical. Here, we extended this recent result to higher $p$-form fields, while also showing that the details of the process can be different. We generalized the simple quartic potential that has been studied for 1-forms to arbitrary $p$-forms, and derived the so-called effective metric that controls the dynamics. If this effective metric becomes non-Lorentzian, one can conclude that the equations of motion do not represent time evolution anymore.

The technical aspects of the problem, as well as the physical relevance led us to consider four spacetime dimensions, where a 2-form field has important differences from 1-forms. For example, the sign of the determinant of the effective metric, which is the primary indicator of hyperbolicity for 1-forms, does not provide sufficient information for 2-forms. Nevertheless, a more detailed analysis concludes that time evolution still breaks down in this case.

Another important difference of higher forms is that they allow for more self interaction terms compared to the 1-form. Restricting ourselves to parity symmetry, 2-forms allow for two different quartic self interaction potentials, both of which lead to loss of hyperbolicity. However, in a surprising twist, these terms can be combined in a way that mitigates hyperbolicity problems. A compound potential improves hyperbolicity by enlarging the part of the configuration space where the theory is healthy, even though there are still unhealthy parts. This is a valuable contribution to the recent efforts to ameliorate hyperbolicity problems in other families of theories~\cite{Thaalba2024}. Our results can be useful in the search for viable alternative ideas in gravity and cosmology by showing which theories are free of undesirable behavior. 

We had the approach that loss of hyperbolicity renders a theory unphysical, but note that this is only the case if we take the equations to be fundamental. If we are considering our action as an \emph{effective} one that is only valid at low energies, then we might rightfully ignore loss of hyperbolicity if it occurs above our energy scale. Then, the resolution of the problem can be delegated to an unknown ultraviolet-complete theory for which our effective field theory (EFT) is only an approximation~\cite{Coates2023b}. Nevertheless, our findings give us clues about which theories are truly fundamental and which can at best be approximations. Moreover, our results on the weakening of loss of hyperbolicity can be useful in obtaining better EFTs.

Finally, the results of this study are generalizable to higher spacetime dimensions. In particular, our conclusions about 3-forms most likely generalize to $(D{-}1)$-forms. Moreover, for even $D$, we can expect $(D/2)$-forms to have a rich structure similarly to 2-forms in four dimensions, since their dual is also a $(D/2)$-form---with even more variety for self interaction.

\acknowledgments
We thank Bayram Tekin and Mehmet Ozkan for their valuable suggestions on the $p$-form field literature. This study was supported by Scientific and Technological Research Council of Turkey (T\"UB\.ITAK) under the Grant Number 122F097. Authors thank T\"UB\.ITAK for its support. 

\appendix
\section{Coordinate singularities and loss of hyperbolicity for $\lambda>0$}
\label{apdx: coordinate singularity}

In the main text, we discussed the time evolution problems that occur due to the breakdown of the physical theory itself. However, one can also encounter problems due to an inadequate choice of formulation for the theory. This happens, for example, when the coordinates that are used to formulate the partial differential equations are chosen without sufficient care. Such \emph{coordinate singularities} are not necessarily a threat to the physical validity of the theory, and they can be, in principle, removed by a change of coordinates, yet, their identification can be quite hard. These issues have been studied for 1-form fields \cite{Coates2022, Coates2023a}, and we will briefly discuss them in relation to 2-forms in the following.

In order to evolve the 2-form equation of motion system in Eq.~\eqref{eq: time evolution system}, the matrix $\mathbb{A}$ needs to be invertible. This yields the condition
\begin{equation}
\det \mathbb{A} = - z^2 \gbar_{tt} \neq 0.
\end{equation}
Note that the case $z=0$, i.e. $x=-1$, cannot be encountered before the actual singularity, since $x_+ \geq -1/3$. However, it is possible that $\gbar_{tt}=0$ is encountered during the evolution. If this happens, the system \eqref{eq: time evolution system} cannot be evolved further in these coordinates. Nevertheless, this does not imply a physical breakdown of the theory, since a foliation of spacetime can be found where $\gbar_{tt}=0$ is not encountered. The close analog of this problem has been studied in detail for 1-forms \cite{Coates2022, Coates2023a}.

Note that $\gbar_{tt} = 0$ occurs when
\begin{equation}
x = -1 + 2\lambda \mathbf{E}^2.
\end{equation}
Thus, if $\lambda<0$, the physical loss of hyperbolicity at $\gbar=0$ always happens before the coordinate singularity, i.e., the coordinate singularity cannot be observed during the evolution. For $\lambda>0$, however, we shall show below that the opposite is true, and the coordinate singularity appears before loss of hyperbolicity. At best, the two singularities coincide, i.e., $x_+ = -1+2\lambda\mathbf{E}^2$, which can be equivalently written as
\begin{equation}
3\lambda\mathbf{E}^2 = 1 + \lambda\mathbf{B}^2 \cos^2\phi,
\end{equation}
where $\phi$ is the angle between $\mathbf{E}$ and $\mathbf{B}$.

To prove the above statements, let us assume that $\lambda>0$, and loss of hyperbolicity has just occurred at $x=x_+$. We shall examine whether it is possible that the coordinate singularity has not been encountered yet, i.e., whether we can have
\begin{equation}
\label{eq: desired order of singularities}
-1 + 2\lambda\mathbf{E}^2 < x_+.
\end{equation}
Note that for given $\mathbf{E}^2$ and $\mathbf{B}^2$, $x_+$ has its maximum when $\mathbf{E}$ and $\mathbf{B}$ are in the same direction; $\cos^2\phi=1$. Thus, our best chance at satisfying Eq.~\eqref{eq: desired order of singularities} is if $\cos^2\phi=1$. In other words, if Eq.~\eqref{eq: desired order of singularities} is satisfied for some $\phi$, then it is satisfied for $\phi=0$ as well. We shall therefore assume $\phi=0$ in the following, and show that the strict inequality \eqref{eq: desired order of singularities} cannot be achieved.

For positive $\lambda$, the condition $x=x_+$ implies $\mathbf{E}\neq0$. We can therefore write
\begin{equation}
\mathbf{B}^2 = \eta \mathbf{E}^2
\end{equation}
for some $\eta\geq0$. Then, the same condition $x=x_+$ gives $3x^2 + 4x + 1 = 4\eta \lambda^2 (\mathbf{E}^2)^2$, or
\begin{equation}
\label{eq: quadratic eqn for lambda E^2}
\left(3\eta-1\right) \left(\eta-3\right) \left(\lambda\mathbf{E}^2\right)^2 + 4\left(\eta-1\right) \lambda\mathbf{E}^2 + 1 = 0.
\end{equation}
This does not have a solution for $\eta=3$. For $\eta=1/3$, the inequality \eqref{eq: desired order of singularities} is not satisfied strictly, but the two singularities coincide; $-1+2\lambda\mathbf{E}^2=x_+$. Excepting these two cases, the solutions of Eq.~\eqref{eq: quadratic eqn for lambda E^2} for $\lambda\mathbf{E}^2$ are
\begin{equation}
\lambda\mathbf{E}^2 = \frac{1}{1-3\eta} \quad \text{and} \quad \lambda\mathbf{E}^2 = \frac{1}{3-\eta}.
\end{equation}
The former can be a solution only for $\eta<1/3$, in which case one cannot satisfy $x=x_+$, since one gets $x<-2/3$. The latter gives coinciding singularities. This completes our proof.

The above means that, in order to check whether dynamical loss of hyperbolicity can occur for positive $\lambda$, we have to devise a time evolution scheme where the coordinate singularity appears simultaneously, which is the best case scenario. This, however, is not straightforward. As a first trial, one can again look at the purely electric or magnetic cases. For $\lambda>0$, we cannot get close to the singularity with a purely magnetic configuration, which is why we shall try a purely electric one.

Let us assume that $\mathbf{B}=0$, and also $\lambda\mathbf{E}^2 \lesssim 1/3$ so that we are close to the loss of hyperbolicity but not quite there. Then, the evolution equations \eqref{eq: time evolution system} tell us that $\mathbf{E}$ stays constant, but a nonzero magnetic part $\delta\mathbf{B}$ develops in general. Expanding to linear order in $\epsilon = \delta\mathbf{B}^2$, we have
\begin{align}
f(x,y) &= \left(3\lambda\mathbf{E}^2 - 4\right) \lambda\mathbf{E}^2 + 1 \nonumber\\
& \quad\quad + 2\lambda \left[2 - \left(3 + 2\cos^2\phi\right) \lambda\mathbf{E}^2 \right] \epsilon.
\end{align}
In the limit $\lambda\mathbf{E}^2 \to 1/3$ this gives
\begin{equation}
f(x,y) = \frac{2\lambda}{3} \left(1 + 2\sin^2\phi\right) \epsilon,
\end{equation}
which is positive. Thus, the evolution of a purely electric configuration close to the singularity seems to bring us farther from it and not nearer, even in the best case scenario $\phi=0$.

Because of this, to show that hyperbolicity can be lost dynamically for $\lambda>0$ as well, one perhaps needs to consider configurations with nonzero electric and magnetic parts that solve the constraint Eq.~\eqref{eq: constraint eqn}. Alternatively, one might consider nontrivial foliations of the underlying Lorentzian spacetime. Such a formulation brings in additional terms to the time evolution, and was successfully used for 1-forms~\cite{Coates2022}. We will not attempt either of these rather-involved strategies here.

\section{Dynamics of 3-form fields}
\label{apdx: 3-form dynamics}

We mentioned in Sec.~\ref{sec: loss of hyperbolicity 3-forms} that free, massless 3-form fields ($m^2=0=V(B)$) are not dynamical in four dimensions, however, this is no longer the case once the original gauge freedom is broken~\cite{Aurilia1981, Aurilia2004, Nitta2018}. In particular, the massive 3-form carries one physical degree of freedom \cite{Aurilia1981}. To demonstrate these points, we will be mainly utilizing the fact that the Hodge dual of a 3-form field is the relatively simpler 1-form field.

Via the duality, it is trivial to see that when we express the mass term in the action of a 3-form field in terms of the dual 1-form, the expression is simply the mass term for a 1-form field, up to a multiplicative constant. The same is also true for the potential terms. However, the canonical kinetic term of the 3-form field changes its nature when expressed in terms of the dual field. That is, in terms of $\tilde{B}_\nu = (1/3!) \epsilon_{\alpha\beta\mu\nu} B^{\alpha\beta\mu}$, the canonical 3-form kinetic term in Eq.~\eqref{eq: Lagrangian of p-form} becomes
\begin{equation}
\label{eq: 3-form kinetic term in terms of dual}
- \frac{1}{2\cdot4!} M^{\mu\nu\alpha\beta}M_{\mu\nu\alpha\beta} = \frac{1}{2} \left(\partial_\mu \tilde{B}^\mu\right)^2.
\end{equation}
On the right hand side is a kinetic term for a 1-form field, but it is \emph{not} the canonical Maxwell kinetic term $F_{\mu\nu}F^{\mu\nu}$.\footnote{The gauge transformation of a $p$-form gauge field $B$ is given, in differential form notation, by $B \to B + \d\xi$, where $\xi$ is an arbitrary ($p{-}1$)-form. This leaves the field strength $M=\d B$ unchanged, and is a simple generalization of the gauge transformations of the Maxwell vector potential. When written in terms of the dual 1-form of a 3-form, however, the same transformation reads as $\tilde{B}_\mu \to \tilde{B}_\mu + \partial^\nu \tilde{\xi}_{\nu\mu}$. Hence, when this theory is viewed as a vector theory, its gauge transformations are different from that of the Maxwell vector potential. Here, the vector changes by the divergence of a 2-form---which leaves the Lagrangian \eqref{eq: 3-form kinetic term in terms of dual} unchanged---and not by the gradient of a scalar.} What is the nature of this 1-form field theory? It is important to understand this point since this 1-form theory is related to the canonical 3-form theory on the left hand side of Eq.~\eqref{eq: 3-form kinetic term in terms of dual}.

Starting from scratch, there are---up to integration by parts---two distinct kinetic terms that can be constructed out of a 1-form $A_\mu$ which are also quadratic in first derivatives; $\partial_\mu A_\nu \partial^\mu A^\nu$ and $(\partial_\mu A^\mu)^2$. \emph{When only a kinetic term is present}, the Maxwell term $F_{\mu\nu}F^{\mu\nu}$ is the only combination of these two that does not lead to ghost degrees of freedom \cite{Rham2014}. This means, strictly speaking, the theory with only Eq.~\eqref{eq: 3-form kinetic term in terms of dual} in its action is unphysical due to ghosts. On the other hand, this theory possesses no physical degrees of freedom. Thus, the existence of the ghost is a nuisance, but also ultimately irrelevant unless we couple the 1-form to other fields.

Addition of mass or self interaction potentials changes the above picture drastically, and makes this theory a healthy one. Via Eq.~\eqref{eq: 3-form kinetic term in terms of dual}, a linear massive 3-form theory with the canonical kinetic term [$m^2 \neq 0, V(B)=0$ in Eq.~\eqref{eq: Lagrangian of p-form}] is also similarly healthy. The equation of motion of this free, massive 3-form can be found from Eq.~\eqref{eq: p-form eom for simplest quartic potential}, or, equivalently from Eqs.~\eqref{eq: p-form Lorenz condition} and \eqref{eq: p-form eom wavelike form} together, by setting $\lambda=0$. Continuing our use of the dual, these equations are perhaps more illustrative when written in terms of $\tilde{B}_\mu$:
\begin{equation}
\square \tilde{B}_\mu = m^2 \tilde{B}_\mu \quad \text{and} \quad \partial_{[\mu} \tilde{B}_{\nu]}=0.
\end{equation}
The latter of these tells us that $\tilde{B}_\mu$ is pure gradient; $\tilde{B}_\mu = \partial_\mu \Phi$, so that the former is equivalent to
\begin{equation}
\left(\square - m^2\right) \Phi = c,
\end{equation}
where $c$ is a constant. Shifting $\Phi$ by a constant as $\Phi \to \Phi - c/m^2$, this is simply the homogeneous Klein-Gordon equation. Hence, the free massive 3-form is really the Klein-Gordon field in disguise \cite{Aurilia1981}. In particular, there are no ghosts or tachyonic instabilities in the theory.

\section{Parity violating interactions}
\label{apdx: parity violating interactions}

We restricted our study to parity symmetric theories in the main text, however, there are also parity-violating terms we can add to the action of a 2-form field.

Even before considering the quartic self interactions, parity-violation can appear at the quadratic order. For $D=4$ consider the action term
\begin{equation}
\frac{\alpha}{8} \epsilon^{\mu\nu\rho\sigma} B_{\mu\nu} B_{\rho\sigma} = \frac{\alpha}{4} B_{\mu\nu} \tilde{B}^{\mu\nu},
\end{equation}
where $\alpha$ is a coupling constant and $\tilde{B}_{\mu\nu}$ is the Hodge dual of $B_{\mu\nu}$. Since this is only quadratic, it is akin to a mass term, whose effect is to replace the constant mass parameter $m^2$ by a (parity-violating) constant \emph{mass tensor}. If we use this term instead of the standard mass term, the field equations become
\begin{equation}
\label{eq: eom for only parity violating mass}
\partial_\alpha M^{\alpha\mu\nu} = \frac{1}{2} \alpha \epsilon^{\mu\nu}_{\phantom{\mu\nu}\alpha\beta} B^{\alpha\beta}.
\end{equation}
However, this theory does not possess any nontrivial solutions. This can be seen easily by applying $\partial_\mu$ to Eq.~\eqref{eq: eom for only parity violating mass}, which implies $M_{\alpha\mu\nu}=0$, which in turn yields $B_{\mu\nu}=0$.\footnote{We could not find a direct reference for this result, but it seems to be simple enough that we believe it to be already known in the physics community.}

This situation changes if both mass terms are present, i.e. consider the Lagrangian \eqref{eq: Lagrangian of p-form} with the potential
\begin{equation}
V = \frac{\beta m^2}{4} B_{\mu\nu} \tilde{B}^{\mu\nu}.
\end{equation}
This yields the equations of motion
\begin{equation}
\partial_\alpha M^{\alpha\mu\nu} = \frac{1}{2} \mathcal{M}^{\mu\nu}_{\phantom{\mu\nu}\alpha\beta} B^{\alpha\beta},
\end{equation}
with
\begin{equation}
\mathcal{M}^{\mu\nu}_{\phantom{\mu\nu}\alpha\beta} = m^2 \left(2\delta^\mu_{[\alpha}\delta^\nu_{\beta]} + \beta \epsilon^{\mu\nu}_{\phantom{\mu\nu}\alpha\beta} \right)
\end{equation}
acting as a mass tensor.

Since this is a linear theory, it can be solved in terms of plane waves. \emph{Going to the rest frame of the plane wave for simplicity}, the general solution can be written as 
\begin{equation}
B_{\mu\nu} = \chi_{\mu\nu}^- e^{-i \omega_m t} + \chi_{\mu\nu}^+ e^{i \omega_m t},
\end{equation}
where $\omega_m \equiv m (1+\beta^2)^{1/2}$ and $\chi^\pm_{\mu\nu}$ are constant with electric and magnetic parts being related by $\mathbf{E}_{\chi^\pm} = \beta\mathbf{B}_{\chi^\pm}$. The dispersion relation can be written in a frame independent way as
\begin{equation}
k^\mu k_\mu + m^2 \left(1+\beta^2\right) = 0.
\end{equation}
In particular, the equations of motion are hyperbolic irrespectively of the value of the coupling constant $\beta$.

Let us now turn to potentials beyond the quadratic order. At first it seems that there are two quartic contractions possible: $\tilde{B}_{\alpha\beta} B^{\alpha\beta} B^2$ and $\tilde{B}_{\alpha\beta} B^{\beta\mu} B_{\mu\nu} B^{\nu\alpha}$, but these are equivalent in four dimensions. Hence, without loss of generality, let us consider the potential
\begin{equation}
V = \frac{\beta m^2}{8} \tilde{B}_{\alpha\beta} B^{\alpha\beta} B^2.
\end{equation}
This gives the equation of motion
\begin{equation}
\partial_\alpha M^{\alpha\mu\nu} = m^2 \tilde{z} B^{\mu\nu} + \beta m^2 \left(\frac{1}{2} B^2\right) \tilde{B}^{\mu\nu},
\end{equation}
where $\tilde{z} = 1 + (\beta/2) \tilde{B}_{\alpha\beta} B^{\alpha\beta}$. Repeating the same steps as before, we get the principal symbol
\begin{align}
\mathcal{P}^{\mu\nu\alpha\beta} &= k^2 g^{\mu\alpha}g^{\nu\beta} + \frac{\beta}{\tilde{z}} k^\mu \Big[2 k_\sigma \tilde{B}^{\alpha\beta} B^{\sigma\nu} \nonumber\\
&\quad\quad + 2B^{\alpha\beta} k_\sigma \tilde{B}^{\sigma\nu} + \frac{1}{2} B^2 \epsilon^{\gamma\nu\alpha\beta} k_\gamma \Big] - \left(\mu\leftrightarrow\nu\right).
\end{align}
Calculating its determinant, we see that it is of the same form as Eq.~\eqref{eq: det of 2-form principal symbol}, for a different $Q_{\alpha\beta\mu\nu}$. We shall therefore pick a frame and focus on a special case again. Let $(k^\mu) = (\omega,0,0,k_z)$ and let $\mathbf{E}$ and $\mathbf{B}$ have the same transverse polarization, e.g., $\mathbf{E} = (E_x,0,0)$ and $\mathbf{B} = (B_x,0,0)$. Then, the condition $Q_{\alpha\beta\mu\nu} k^\alpha k^\beta k^\mu k^\nu = 0$ yields
\begin{equation}
\label{eq: dispersion relation pv quartic special case}
\omega^4 - 2\left[\frac{2\beta^2}{a}\left(E_x^2 + B_x^2\right)^2 + 1\right] k_z^2 \omega^2 + k_z^4 = 0,
\end{equation}
where
\begin{equation}
a = \left(1 + 2\beta E_x B_x\right) \left(1 + 6\beta E_x B_x\right).
\end{equation}
It can be seen that if the field grows past $6\beta E_x B_x = -1$, this dispersion relation will become problematic. For instance, for $E_x = -B_x = (3\beta)^{-1/2}$ the dispersion relation \eqref{eq: dispersion relation pv quartic special case} reads as
\begin{equation}
\left(3\omega^2 + k_z^2\right) \left(\omega^2 + 3k_z^2\right) = 0,
\end{equation}
which has no real solutions. We therefore see that the configuration space includes pathological regions.

\end{document}